\documentclass[prl,twocolumn,superscriptaddress]{revtex4}
\usepackage{amsmath,amssymb,graphicx}

\begin{document}
\title{Polaritonic analogue of Datta and Das spin transistor}

\author{R. Johne}
\affiliation{LASMEA, CNRS-Universit\'e Blaise Pascal, 24 Avenue des Landais, 63177
Aubi\`ere Cedex France}

\author{I. A. Shelykh}
 \affiliation{Science Institute, University of Iceland, Dunhagi-3, IS-107, Reykjavik, Iceland}
 \affiliation{St. Petersburg State Polytechnical University,
 195251, St. Petersburg, Russia}

\author{D. D. Solnyshkov}
\affiliation{LASMEA, CNRS-Universit\'e Blaise Pascal, 24 Avenue des Landais, 63177
Aubi\`ere Cedex France}

\author{G. Malpuech}
\affiliation{LASMEA, CNRS-Universit\'e Blaise Pascal, 24
Avenue des Landais, 63177 Aubi\`ere Cedex France}

\date{\today}

\begin{abstract}
We propose the scheme of a novel spin-optronic device, optical
analog of Datta and Das spin transistor for the electrons. The role
of the ferromagnetic--nonmagnetic contact is played by a spatially
confined cavity polariton BEC. The
condensate is responsible for the appearance of effective magnetic
field which rotates the spin state of a propagating pulse of
polaritons allowing to tune the transmittivity of the device.

\end{abstract}

\maketitle

Spintronics is one of the trends in modern mesoscopic physics
\cite{Zutic2004}. It was born in 1990, when S. Datta and B. Das in their
pioneer work proposed a theoretical scheme of the first spintronic
device \cite{DattaDas}, which afterwards was named Datta and Das
spin transistor. It consists of two ferromagnetic 1D or 2D
electrodes, usually with collinear magnetizations, separated by a
non-magnetic semiconductor region in which a Spin-Orbit Interaction
(SOI) of the Rashba type is induced by a top gate electrode,
\begin{equation}
\widehat{H}_{SOI}=\alpha
\left[\widehat{\textbf{k}}\times\sigma\right]\cdot\textbf{e}_z,
\label{RashbaSOI}
\end{equation}
where $\textbf{e}_z$ is a unity vector in the direction of the
structure growth axis $z$, $\sigma$ denotes a set of Pauli matrices,
$\widehat{\textbf{k}}=-i\nabla$. $\alpha$ is a characteristic Rashba
parameter, which depends on the degree of asymmetry of a quantum
well (QW) in the $z$ direction. It can be efficiently tuned by
varying the top gate voltage $V_g$
\cite{Nitta1997,Heida1998,Engels1997}. The Hamiltonian can be
interpreted in terms of an effective magnetic field lying in the
plane of a QW and being perpendicular to the carriers' kinetic
momentum. This effective field provokes the rotation of the spin of
the carriers in the semiconductor region and results in the
oscillations of the transmitted current $I_{tr}$ as a function of
Rashba coupling controlled by the gate voltage $V_g$: $ I_{tr}\sim
\textrm{cos}^2\left(2m_{\text{eff}}\alpha L/\hbar^2\right),$ where
$m_{\text{eff}}$ is the carrier effective mass in the semiconductor.
The above formula has a very clear physical meaning: only the spin
component parallel to the magnetization can propagate in the
outgoing ferromagnetic lead. The outgoing current should be thus
dependent on the rotation angle accumulated over a distance L
between the two leads, $\Delta\phi=2m_{\text{eff}}\alpha L/\hbar^2$.

Although the scheme of the Datta and Das spin transistor seems very
simple from a theoretical point of view, its practical
implementation appears to be extremely complicated due to the
problems of spin injection, decoherence and realisation of an abrupt enough
ferromagnetic-non magnetic junction. These difficulties led to alternative
propositions of spintronic devices which do not need spin injection
\cite{Aronov1993,Konig2006}. More than 15 years of intensive
experimental work in this direction did not result in any
breakthrough and the Datta and Das device still remains a
theoretical concept \cite{Bratkovsky2008}.

On the other hand, it was recently proposed that in the domain of
mesoscopic optics the controllable manipulation of the (pseudo)spin
of excitons and exciton-polaritons (polaritons) can provide a basis for the
construction of optoelectronic devices of the new generation, called
spin-optronic devices,  that would be the optical analogs of
spintronic devices. The first element of this type, namely
polarisation-controlled optical gate, was recently realised
experimentally \cite{Leyder2007}, and the principal schemes of other
devices such as Berry phase interferometer \cite{Shelykh2009}, have
been proposed theoretically. It has also been demonstrated by several experimental groups
that equilibrium polariton Bose Einstein condensation (BEC) can be achieved \cite{Kasprzak,Deng2006, Baumberg2008,Kasprzak2008,Wertz2009}.
Also the spatial modulation of the polariton
wavefunction and polariton condensates is now well controlled
experimentally \cite{Balili2007,Lai2007,Sanvitto2009,Cerna2009},
offering extremely wide perspectives for the implementation of
polaritons circuits.

Polaritons are the elementary
excitations of semiconductor microcavities in the strong coupling
regime. An important peculiarity of the polariton system is its spin
structure: being formed by bright heavy-hole excitons, the lowest
energy polariton state has two allowed spin projections on the
structure growth axis ($\pm1$), corresponding to the right and left
circular polarisations of the counterpart photons. The states having
other spin projections are split-off in energy and normally can be
neglected while considering polariton dynamics. Thus, from the
formal point of view, the spin structure of cavity polaritons is
similar to the spin structure of electrons (both are two-level
systems), and their theoretical description can be carried out along
similar lines. The possibility to control the spin of
cavity polaritons opens a way to control the polarisation of the light
emitted by a cavity, which can be of importance in various
technological implementations including optical information
transfer.

It should be noted, however, that the fundamental nature of
elementary excitations is different in two kinds of systems:
electrons and holes (i.e. fermions) in the case of spintronics,
polaritons (i.e. bosons) in the case of spin-optronics.
Also, it appears that the account of many-body interactions is of
far greater importance for spinoptronic devices with respect to the
spintronic ones. The polariton-polariton interactions in
microcavities are strongly spin-anisotropic: the interaction of
polaritons in the triplet configuration (parallel spin projections
on the structure growth axis) is much stronger than that of
polaritons in the singlet configuration (antiparallel spin
projections)\cite{Combescot2006}. This leads to a mixing of linearly
polarised polariton states which manifests itself in remarkable
nonlinear effects, which are of great importance for the functioning
of spinoptronic devices in nonlinear regime.

As shown in \cite{Shelykh2009}, the analogue of Rashba SOI in
microcavities can be provided by the longitudinal-transverse
splitting (TE-TM splitting) of the polariton mode. However, the
TE-TM splitting cannot be easily tuned by the simple application of
a voltage, unlike the Rashba SOI. In ring interferometers the
control of the polariton Berry phase which governs the interference
pattern therefore requires to modulate an external magnetic field,
which is expected to be relatively slow. In the present paper we
propose a completely new and tunable way to realise an optical
ferromagnetic-nonmagnetic contact, which will finally allow to
design a nano-device, optical analogue of the Datta and Das spin
transistor.

\begin{figure}
\begin{center}
\includegraphics[width=1.0\linewidth]{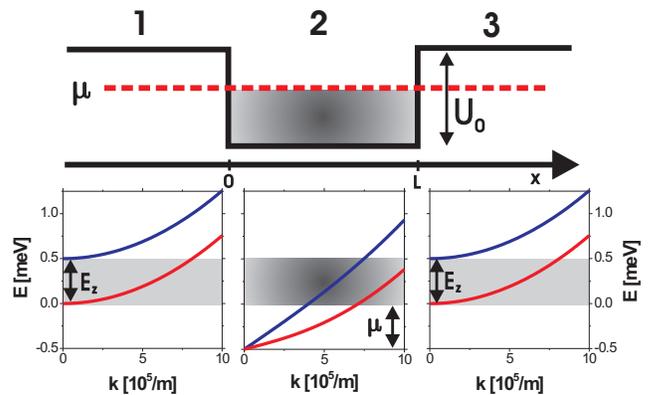}
\caption {\label{fig1} Schematic illustration of a polariton spin transistor (upper panel) with the corresponding renormalised polariton dispersions (lower panel). The regions are indicated with 1 ($x<0$), 2 ($0<x<L$) and 3 ($x>L$). Dark gray (blue) corresponds to $\sigma_-$ and light gray (red) for $\sigma_+$ polarisation. In region 2 the mixing of both circular polarisations results in an elliptical polarised condensate, while in flanking regions only the $\sigma_+$ component can propagate.}
\end{center}
\end{figure}

We propose to use the change of the polarisation eigenstates induced by the
formation of polariton BEC in the
presence of magnetic field as the analogue of the Rashba field. In
that case the modulation of the signal will be driven not by the
modulation of the magnetic field, but by the modulation of the
condensate density, which can be achieved either by the modulation
of a pumping laser intensity, or by the modulation of a voltage in
case of electrically pumped condensate
\cite{Bajoni2007,Khalifa2008,Tsintzos2008}. The device is
constituted by a planar microcavity showing a confining potential
having the shape of a stripe of width L as shown on the upper panel
of the Fig.(\ref{fig1}).

We divide the system into three regions: (1) $x<0$, (2) $0<x<L$ and
(3) $x>L$. We assume that critical conditions for the formation of a
quasi-equilibrium BEC of polaritons are fulfilled as demonstrated experimentally by several
independent groups \cite{Kasprzak,Deng2006,Kasprzak2008,Wertz2009}.
We also assume that the chemical potential $\mu$ stands below the
edge of the barriers, so that the condensate is confined in the
central region and absent in the flanking regions.
We consider the effect of an external magnetic field $B$ applied
perpendicularly to the structure interface. In the lateral regions 1
and 3, the normal Zeeman splitting $E_z$ between the polariton modes
occurs as shown in the lower panel of the fig 1. This opens an
energy gap $E<E_z=\mu_b g B$ where only one of the two circularly
polarised component can propagate. We assume here and in the
following that the Zeeman splitting is much larger than the TE-TM
splitting which can therefore be neglected. In the central region,
however, the presence of the condensate leads to the full
paramegnetic screening, also known as spin-Meissner effect
\cite{SpinMeissner}. For a given field $B$, the critical density
$n_c$ in the polariton condensate can be defined as $n_c=\mu_b g
B/(\alpha_1-\alpha_2),$ where $\alpha_{1(2)}$ are the interaction
constants for particles with the same (perpendicular) spin
projection, $g$ is the exciton g-factor and $\mu_b$ is the Bohr
magneton. Below this critical density $n_c$, the spin anisotropy of
the polariton-polariton interactions leads to a full paramagnetic
screening of the Zeeman splitting $E_Z$  resulting in a quenching of
the Zeeman gap, as shown in the lower part of the Fig.\ref{fig1}.
The polariton condensate is elliptically polarised, which is also
the case for the propagative modes in the central region. The
polarisation degree of these modes depends on the condensate
density. Therefore a circularly polarised $\sigma_+$ pulse with an
energy located within the Zeeman gap of the lateral regions can
enter into the central region. During its propagation in this region
its polarisation vector will be rotated by an effective magnetic
field whose direction is associated with the polarisation of the
eigenstates in this region. This effective "spin-Meissner field" has
some in-plane component and plays the role of the Rashba SOI
effective field. The intensity of the outgoing current depends on
the angle $\Delta\phi$ between the pseudospin vector of the
polaritons reaching the outgoing lead. If the precession is such
that the pulse becomes fully $\sigma_-$ polarised on the interface
between 2 and 3, the pulse will be fully reflected. If the pulse is
fully $\sigma_+$ polarised, it will be fully transmitted. Working in
this energy range means that for polaritons we create a situation
analogical to ferromagnetic-nonmagnetic-ferromagnetic interface,
which one needs for a creation of the Datta and Das device.

Such a configuration has a number of possible advantages with
respect to classical spintronics: the dramatic impact of carrier
spin relaxation or decoherence, which has severely limited the
achievement or the functionality of any semiconductor-based
spintronic devices, is strongly reduced \cite{Langbein2007}. Besides, the
solution of the spin injection problem is now trivial: it is performed
simply by choosing an appropriate polarisation of the exciting
laser.

Quantitatively, the outgoing amplitude can be calculated by solving a system of linear equations.
The wavefunction of a propagating mode in the three regions can be written in the following way:
\begin{eqnarray}
\label{wavefunction1}
\Psi_1=&(e^{ik x}+ r e^{ik x}) \begin{pmatrix} 1 \\ 0 \end{pmatrix} + A e^{\gamma x} \begin{pmatrix} 0 \\ 1 \end{pmatrix}, \\
\label{wavefunction2} \Psi_2=&(C_1^+ e^{ik_1 x}+ C_{1}^- e^{-ik_1
x}) \begin{pmatrix} \cos \beta \\ \sin \beta \end{pmatrix}+ \\
\nonumber
&+(C_{2}^+ e^{ik_2 x}+ C_2^- e^{-ik_2 x}) \begin{pmatrix} -\sin \beta \\ \cos \beta \end{pmatrix}, \\
\label{wavefunction3} \Psi_3=&(te^{ik x}) \begin{pmatrix} 1 \\ 0
\end{pmatrix} + D e^{-\gamma x}\begin{pmatrix} 0 \\ 1 \end{pmatrix},
\end{eqnarray}
where $r$ is the amplitude of reflectivity, $t$ is is transmission amplitude, $C_{1,2}^{+(-)}$ are the complex amplitudes of forward (backward) running waves in the trap with different polarisations and wavevectors $k_1$ and $k_2$. The wavevectors are determined by the dispersion relations for each region Ref.\cite{SpinMeissner}, which read:
\begin{eqnarray}
k=&\sqrt{\frac{2m}{\hbar^2}E}, \nonumber
\gamma=\sqrt{\frac{2m}{\hbar^2}\left(E_z-E\right)} \\ \nonumber
k_{1,2}=&\sqrt{ \frac{m}{\hbar^2}(-n_2
U_{1,2}+\sqrt{4(E-\mu)^2+n_2^2 U_{1,2}^2})} \\ \nonumber
U_{1,2}=&\alpha_1 \pm\sqrt{\alpha_2^2+(\alpha_1^2-\alpha_2^2)\left(B/B_{C}\right)^2}.
\end{eqnarray}
The polarisation of the excitations in the regions 1 and 3 is $\sigma_+$ and $\sigma_-$. The polarisation of the elementary excitations of the condensate in the spin-Meissner phase (region 2) has never been calculated. It can be found by the standard method of linearisation with
respect to the amplitude of the elementary excitations of the condensate, which gives the following result:
\begin{equation}
\label{beta} \tan\beta=\frac{-\alpha_1 \cos2\Theta +\sqrt{\alpha_1^2
\cos^2 2\Theta+\alpha_2^2 \sin^2 2\Theta} }{\alpha_2 \sin 2\Theta},
\end{equation}
where $\Theta= \frac{1}{2} \arcsin\sqrt{1-\left(B/B_C\right)^2}$.

Interestingly, the polarisation of the excitations, associated with
the angle $\Theta$, is different from the one of the condensate,
associated with the angle $\beta$. The precession frequency in the
spin-Meissner effective field directed along $(\cos\beta,\sin\beta)$,
can be estimated as
\begin{equation}
\Omega_t = \frac{E}{{2\hbar }}\frac{{k_1 ^2  - k_2 ^2 }}{{k_1  k_2  }}.
\end{equation}
To find the amplitude of the outgoing beam one has different possibilities.
Using an analytical approach similar to the transfer matrix method, one obtains for the reflected $r$ and transmitted $t$ amplitudes:
\begin{eqnarray}
&&t = e^{ik_1  L} \cos ^2 \beta + e^{ik_2  L} \sin ^2 \beta+
\\ \nonumber && + \frac{{\left[ {\cos \beta\sin \beta
\left( {e^{ik_1  L}  - e^{ik_2  L} } \right)} \right]^2 \left[ {\sin
^2 \beta e^{ik_1  L}  + \cos ^2 \beta  e^{ik_2  L} } \right]}}{{1 -
\left[ {\sin ^2 \beta e^{ik_1  L}  + \cos ^2 \beta e^{ik_2 L} }
\right]^2 }} \\ \nonumber &&r = \frac{{\left( {\left[ {e^{ik_1 L}  -
e^{ik_2  L} } \right]\cos \beta \sin \beta} \right)^2 }}{{1 - \left[
{\sin ^2 \beta e^{ik_1  L}  + \cos ^2 \beta e^{ik_2 L} } \right]^2
}}.
\end{eqnarray}
In this analytical approach, the $\sigma_-$ polarised part is assumed to be fully reflected and does not have decaying tails outside the central region. A more exact approach is to use the wavefunctions (Eqs. (\ref{wavefunction1})-(\ref{wavefunction3})) without this approximation, applying corresponding boundary conditions, which ensure the continuity of the wavefunction and current conservation at the interfaces.
\begin{figure}
\begin{center}
\includegraphics[width=0.8\linewidth]{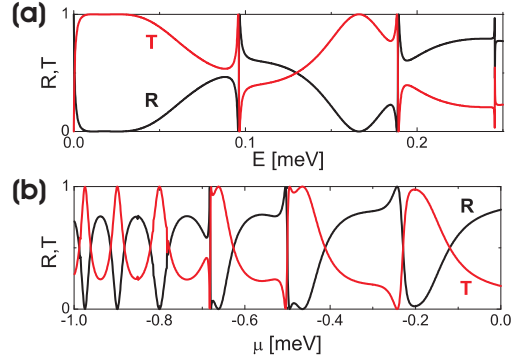}
\caption {\label{fig2} Reflection and transmission coefficient
versus energy of  the probe pulse excitation energy: (a) Dependence
of T (red) and reflection coefficient R (black) on the excitation
energy E ($\mu=-0.7$ meV) and (b) T and R versus chemical potential
$\mu$ ($E=0.2$ meV). The used parameters are: $B=5$ T, $E_Z=0.25$
meV, $L=20 \mu$ m, and $U_0=-1$ meV.}
\end{center}
\end{figure}
Fig.\ref{fig2} shows the dependence of the transmission coefficient $T=|t|^2$ and reflection coefficient $R=|r|^2$ on the excitation energy E (a) and on the chemical potential of the condensate $\mu$ (b). The polarisation of the particles is rotated during the propagation in region 2 with elliptically polarised excited states. The calculation is performed taking into account realistic parameters of a GaAs microcavity (listed in the figure caption). Varying the energy of the injected particles, which can be done by changing the excitation angle of the resonant laser, increases or decreases the value of the spin-Meissner effective field affecting the particle propagation in the central region. Another way to modulate the outgoing beam keeping the particle energy constant (close to the resonances on Fig.\ref{fig2}(a)), is to change the particle concentration (and thus the chemical potential $\mu$) in region 2. This can be simply realised by the modulation of the optical or electrical pumping of the condensate. The impact of the particle concentration is shown in Fig.\ref{fig2}(b). Close to the resonances, the outgoing beam drops from full transmission to zero transmission for a very weak change of $\mu$. This gives a possibility to tune rapidly the outgoing current of the proposed device simply by changing one external parameter.

Of course the outgoing intensity can also be modulated by the magnetic field, but the change of the magentic field intensity is rather slow in comparison to intensity modulation of a pumping laser or to modulation of the applied voltage (in case of electrical pumping).
From the point of view of experimental realisation, increasing the magnetic field enlarges the Zeemann splitting and the energy range where only one polarisation component can propagate outside the condensate. Thus, on one hand, it should be preferable to apply a huge magnetic field in order to increase the operating energy range. On the other hand, high magnetic fields complicate the practical applications.

We have also performed a realistic numerical simulation of the
device operation. We have first calculated the wavefunction of the
condensate in the trap region by minimizing the free energy
 of the system on a grid, taking into account
the two polarisation components and the interactions between them.
All the parameters are taken the same as in Fig.\ref{fig2}.

We have then simulated the propagation of a circularly polarised pulse through the system with the spinor Gross-Pitaevskii equation for polaritons using the wavefunction of the condensate found previously. The main difference with the analytical model presented above is the spatial profile of the wavefunction of the condensate; this difference becomes negligible at larger densities, when the interaction energy is much higher than the kinetic one and the wavefunction achieves the Thomas-Fermi profile. Another difference is that we study the propagation of a pulse, more important from the practical point of view, instead of solving a steady-state problem.
\begin{figure}
\begin{center}
\includegraphics[width=1.0\linewidth]{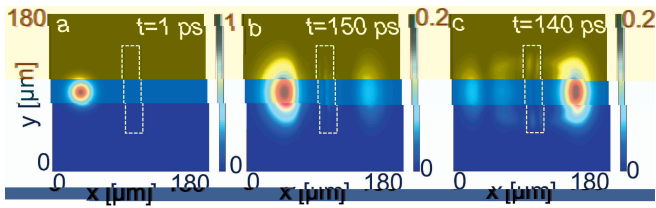}
\caption {\label{GP} Calculated propagation of a wavepacket through the spin transistor: (a) wavepacket created by a laser pulse;  (b) wavepacket reflected by the condensate;  (c) wavepacket transmitted almost without reflection. }
\end{center}
\end{figure}
Figure \ref{GP} shows the snapshots of the pulse propagation through the device. Full movies are available in the Auxiliary Material \cite{EPAPS}. Only the wavefunction of the pulse is shown, without the condensate located in the trap shown by the rectangle. Panel (a) shows the initial state: a Gaussian wavepacket is created by a short laser pulse to the left of the trap with the condensate. Panels (b,c) show the system after the wavepacket has interacted with the condensate: in (b) the packet is mostly reflected, whereas in (c) a larger part passes through. The two latter panels correspond to two different regimes of the spin transistor operation depending on the condensate density: closed (b) and open (c). The broadening of the wavepacket is due to the interaction with the condensate; however this relatively small broadening should not be detrimental for the device.

In conclusion, we proposed a scheme of a polaritonic analogue of
Datta and Das spin transistor. The proposed geometry allows to solve
the problems of decoherence and inefficient spin injection which
were blocking the experimental implementation of Datta and Das spin
transistor for electrons. The requirement of abrupt interfaces seems also easier to achieve
for polaritons which are more extended particles than electrons.
The role of the the non- magnetic region
is played by a confined spinor polariton BEC. The polariton BEC
provokes the appearence of an effective "spin-Meissner" magnetic
field which is acting on the pseudospin of the propagating polaritons.
This field and therefore the device transmissivity is easily and
quickly controlled tuning the condensate density which in an ideal
case can be done changing an applied voltage.

\end{document}